# Combinatorial Algorithms for Capacitated Network Design

MohammadTaghi Hajiaghayi[*]    Rohit Khandekar[†]    Guy Kortsarz[‡]    Zeev Nutov [§]

October 31, 2018

**Abstract**

We focus on designing *combinatorial algorithms* for the CAPACITATED NETWORK DESIGN problem (CAP-SNDP). The CAP-SNDP is the problem of satisfying connectivity requirements when edges have costs and *hard* capacities. We begin by showing that the GROUP STEINER TREE is a special case of CAP-SNDP even when there is connectivity requirement between only one source-sink pair. This implies the *first* poly-logarithmic lower bound for the CAP-SNDP.

We next provide *combinatorial* algorithms for several special cases of this problem. The CAP-SNDP is equivalent to its special case where every edge has either zero cost or infinite capacity. We consider a special case, called CONNECTED CAP-SNDP, where all infinite-capacity edges in the solution are required to form a connected component containing the sinks. This problem is motivated by its similarity to the Connected Facility Location problem [20, 31]. We solve this problem by reducing it to SUBMODULAR TREE COVER, which is a common generalization of CONNECTED CAP-SNDP and GROUP STEINER TREE. We generalize the recursive greedy algorithm [10] achieving a poly-logarithmic approximation algorithm for SUBMODULAR TREE COVER. This result is interesting in its own right and gives the first poly-logarithmic approximation algorithms for Connected hard capacities set multi-cover and Connected source location.

We then study another special case of CAP-SNDP called UNBALANCED-P2P. Besides its practical applications to shift design problems [13], it generalizes many problems such as $k$-MST, Steiner Forest and Point-to-Point Connection. We give a combinatorial logarithmic approximation algorithm for this problem by reducing it to degree-bounded SNDP.

---

[*]University of Maryland, College Park, MD 20742, U.S.A., E-mail: hajiagha@cs.umd.edu. The author is also with AT&T Labs–Research, Florham Park, NJ 07932, U.S.A.

[†]IBM T.J. Watson research center, Hawthorne, NY 10532, U.S.A., E-mail: rohitk@us.ibm.com.

[‡]Rutgers University–Camden, NJ 08102, U.S.A., E-mail: guyk@camden.rutgers.edu. Research partially supported by NSF grant 0819959.

[§]The Open University of Israel. Raanana, Israel. E-mail:nutov@openu.ac.il.

# 1 Introduction

## 1.1 CAPACITATED SURVIVABLE NETWORK DESIGN

The main topic of this paper is the following fundamental network design problem.

> CAPACITATED SURVIVABLE NETWORK DESIGN (CAP-SNDP)
> *Instance:* An undirected graph $G = (V, E)$ with edge-capacities $\{u_e \mid e \in E\}$ and edge-costs $\{c_e \mid e \in E\}$, and requirements $\{r_{ij} \mid i, j \in V\}$ ($c_e$, $u_e$, and $r_{ij}$ are all non-negative integers).
> *Objective:* Find a minimum cost spanning subgraph $H$ of $G$ such that for every $i, j \in V$, the capacity of any $ij$-cut in $H$ is at least $r_{ij}$ (equivalently, $H$ contains $r_{ij}$ $ij$-paths such that every edge $e$ belongs to at most $u_e$ paths).

The special case with a single source and a single sink is called *fixed cost flow* [15]. When all edge-capacities are unit, CAP-SNDP reduces to the Survivable network design problem (SNDP) [18, 23], which generalizes Steiner forest problem [2]) when the connectivity requirements are in $\{0, 1\}$ and Steiner tree problem [15] when connectivity requirements are in $\{0, 1\}$ and all sinks are identical. Unlike these classical special cases, however, the approximability of CAP-SNDP is not well understood; not even a logarithmic hardness is known, and at the same time no better than $o(|E|)$-approximation algorithm is known, even for very restricted settings.

The CAP-SNDP also generalizes the following *buy-at-bulk*-type network design problem. Given an undirected graph $G = (V, E)$ where each edge $e \in E$ is associated with a non-decreasing cost function $f_e$ that specifies the cost $f_e(u)$ of installing $u$ units of capacity on edge $e$. The instance also gives connectivity requirements $\{r_{ij} \mid i, j \in V\}$. The problem is to decide the capacity $u_e$ to be installed on edge $e$ so that the capacity of any $ij$-cut is at least $r_{ij}$ and the total cost $\sum_{e \in E} f_e(u_e)$ is minimized. Note that both the recently studied versions, namely one with economies of scale in which functions $f_e$ are concave [11, 30] and one with dis-economies of scale in which functions $f_e$ are non-concave [3], are special cases of this problem. Assuming that all the involved numbers are polynomially bounded integers, we can reduce this problem to CAP-SNDP by replacing each edge $e$ with parallel edges $e_1, e_2, \ldots, e_R$ where edge $e_k$ has capacity $u_{e_k} = 1$ and cost $c_{e_k} = f_e(k) - f_e(k-1)$. Here $R = \max_{i,j} r_{ij}$. If the numbers are not polynomially bounded, we can use standard scaling techniques to get a polynomial reduction, while losing a constant factor in the approximation guarantee.

It is easy to see that CAP-SNDP is equivalent to its special case where each edge has either infinite (or sufficiently large) capacity or zero cost. Such a reduction is done by replacing each edge $e$ of capacity $u_e$ and cost $c_e$ by a path of length two with two edges $e_1$ and $e_2$ where $u_{e_1} = \infty, c_{e_1} = c_e$ and $u_{e_2} = u_e, c_{e_2} = 0$. We call an edge with infinite capacity a *cost-edge* and an edge with zero cost a *capacity-edge*. From now on, we assume that our CAP-SNDP instances satisfy this property.

## 1.2 CONNECTED SINGLE-SINK CAPACITATED SURVIVABLE NETWORK DESIGN

We next consider the following special case of CAP-SNDP. This special case is motivated by its similarity with Connected Facility Location problem where the open facilities are required to be connected by a backbone network [20, 31].

> CONNECTED SINGLE-SINK CAPACITATED SURVIVABLE NETWORK DESIGN (CONNECTED CAP-SNDP)
> *Instance:* An undirected graph $G = (V, E_u \cup E_c)$ where $E_u$ is the set of capacity-edges (with integer capacities $u_e$ and zero cost) and $E_c$ is the set of cost-edges (with integer costs $c_e$ and infinite capacity), a sink $t \in V$ and sources $s_1, \ldots, s_k \in V$ with their integer requirements $r_1, \ldots, r_k$.
> *Objective:* Find a minimum-cost subgraph $H$ of $G$ such that the minimum $s_i t$-cut in $H$ has capacity at least $r_i$ (for $i \in [k]$) and $H \cap E_c$ forms a connected (backbone) graph containing $t$.

Our first main result is as follows. Let $n$ denote the number of nodes in $G$.

**Theorem 1.1** *Even the single-source version of* CONNECTED CAP-SNDP *is* $\Omega(\log^{2-\epsilon} n)$*-hard to approximate for any* $\epsilon > 0$*, unless NP has a Las Vegas quasi-polynomial-time algorithm. Thus* CAP-SNDP *is also* $\Omega(\log^{2-\epsilon} n)$*-hard to approximate for any* $\epsilon > 0$*, unless NP has a Las Vegas quasi-polynomial-time algorithm. Here* $n = |V|$ *denotes the number of vertices.*

**Theorem 1.2** *There exists a polynomial-time combinatorial* $\Omega(\log^{2+\epsilon} n \cdot \log \sum_{i=1}^{k} r_i)$*-approximation algorithm for* CONNECTED CAP-SNDP *for any* $\epsilon > 0$*.*

We prove Theorem 1.1 by giving an approximation ratio preserving reduction from the well-knwon GROUP STEINER TREE problem to the single-source version of CONNECTED CAP-SNDP. Recall that the GROUP STEINER TREE problem is defined as follows. The instance is given by an undirected graph $G = (V, E)$ with edge-costs $c_e$ and subsets (groups) of nodes $g_1, \ldots, g_k \subseteq V$ and the objective is to find a minimum-cost subtree $H$ of $G$ that contains at least one node from every group. Halperin and Krauthgamer [22] state the lower bound in the form of $\Omega(\log^{2-\epsilon} k)$ where $k$ is the number of groups. However, the size of their construction is $O(k)$. In particular, the number of nodes they have in the tree is less than $2k$. We present the lower bound in terms of the number $n$ of nodes in their tree, which is between $k$ and $2k$. Since the lower bound is polylogarithmic, $k$ and $n$ are essentially the same for our purposes.

## 1.3 SUBMODULAR TREE COVER

We prove Theorem 1.2 by presenting an algorithm for a very interesting generalization of CONNECTED CAP-SNDP called SUBMODULAR TREE COVER. We define SUBMODULAR TREE COVER below and show that it in fact generalizes several other interesting problems. Let $U$ be a ground-set $U$. A function set-function $f : 2^U \to \mathbb{Z}$ is called *non-decreasing* if $f(A) \leq f(B)$ for all $A \subseteq B \subseteq U$, and is called *submodular* if $f(A) + f(B) \geq f(A \cap B) + f(A \cup B)$ for all $A, B \subseteq U$.

SUBMODULAR TREE COVER
*Instance:* An undirected graph $G = (V, E)$ with edge-costs $\{c_e \mid e \in E\}$ and a non-decreasing submodular function $f : 2^V \to \mathbb{Z}$. The function $f$ is by a *value oracle* that returns $f(S)$ when given $S \subseteq V$.
*Objective:* Find a minimum-cost sub-tree $T = (V_T, E_T)$ of $G$ such that $f(V_T) = f(V)$.

We show the following algorithmic result for SUBMODULAR TREE COVER. Let $n$ denote the number of nodes in $G$ and $F_{\max} = f(V) = \max_{S \subseteq V} f(S)$ be the maximum value of $f$. For the purpose of this paper, we assume that $F_{\max}$ is polynomially bounded in $n$.

**Theorem 1.3** *There exists a polynomial-time combinatorial* $O(\log^{2+\epsilon} n \cdot \log F_{\max})$*-approximation algorithm for* SUBMODULAR TREE COVER*, for any* $\epsilon > 0$*.*

We now argue that SUBMODULAR TREE COVER generalizes the following interesting problems.

- CONNECTED CAP-SNDP.
  Given an instance $(G = (V, E_u \cap E_c), t \in V, s_1, \ldots, s_k \in V, r_1, \ldots, r_k \geq 0)$ of CONNECTED CAP-SNDP, we construct an instance of SUBMODULAR TREE COVER as follows. Let $G_c = (V, E_c)$ be the graph on the cost-edges with costs $\{c_e \mid e \in E_c\}$ inherited from the CONNECTED CAP-SNDP instance. Similarly let $G_u = (V, E_u)$ be the graph on the capacity-edges with capacities $\{u_e \mid e \in E_u\}$ inherited from the CONNECTED CAP-SNDP instance. Given $S \subseteq V$ and $i \in [k]$, let $u(\delta(S, s_i))$ denote the capacity of the minimum capacity cut in $G_u$ that separates $s_i$ from all vertices in $S$. Now define a set-function $f : 2^V \to \mathbb{Z}$ as follows. For $S \subseteq V$, let

$$f(S) = \sum_{i=1}^{k} \min\{r_i, u(\delta(S, s_i))\}.$$

It is easy to see that $f$ is non-decreasing. To show that $f$ is submodular, it is enough to argue that $u(\delta(S, s_i))$ is submodular for any $i$. Now for any two sets $S_j \subseteq V$ for $j = 1, 2$, let $S_j \subseteq C_j \not\ni s_i$ be the minimum capacity cuts that separate $s_i$ from $S_j$. Note that $u(\delta(C_1)) + u(\delta(C_2)) \geq u(\delta(C_1 \cap C_2)) + u(\delta(C_1 \cup C_2))$. Since $C_1 \cap C_2$ (resp., $C_1 \cup C_2$) separate $s_i$ from $S_1 \cap S_2$ (resp., $S_1 \cup S_2$), the claim follows. Note also that that there is a polynomial-time value oracle based on minimum cut computations. Theorem 1.3 implies Theorem 1.2.

The *source location problem* studied by Bar-Ilan et al. [4] can be thought of as a special case of CONNECTED CAP-SNDP. Our result gives the first non-trivial approximation algorithm for this problem in the setting of *general graph connectivity cost functions*.

- GROUP STEINER TREE with *group-demands* and *node-capacities*.
  The instance of this problem is the same as for GROUP STEINER TREE except that in addition, there is a *demand* $d_i$ for every group $g_i, i \in [k]$ and a *capacity* $b_v$ for every node $v \in V$. The objective is to compute a minimum-cost sub-tree $T = (V_T, E_T)$ of $G$ and *assign* each node $v \in V_T$ to one or more groups $g_i$ such that

    - if $v \in V_T$ is assigned to a group $g_i, i \in [k]$, then $v \in g_i$,
    - each node $v \in V_T$ is assigned to at most $b_v$ groups, and
    - at least $d_i$ nodes are assigned to each group $g_i, i \in [k]$.

  The fact that this problem is a special case of SUBMODULAR TREE COVER can be shown with the function $f : 2^V \to \mathbb{Z}$ defined below. Fix a subset $S \subseteq V$ and construct a flow network $N$ as follows. The vertices in $N$ are `source`, `sink`, $z_v$ for $S \in V$ and $y_i$ for groups $i \in [k]$. The directed arcs in $N$ are (`source`, $z_v$) with capacity $b_v$ for $v \in S$, ($z_v, y_i$) with capacity 1 for $i \in [k]$ and $v \in g_i \cap S$ and ($g_i$, `sink`) with capacity $d_i$ for $i \in [k]$. Define $f(S)$ to be the maximum flow that can be routed from `source` to `sink` in $N$. It is easy to see that $f$ is a non-decreasing and submodular function. Theorem 1.3 implies $O(\log^{2+\epsilon} n \cdot \log \sum_{i=1}^{k} d_i)$ approximation algorithm for this problem.

  The special case with $b_v = \infty$ for all nodes $v$ is called a Covering Steiner tree problem and is known to admit an $O(\log^3 n)$-approximation ratio [25, 13, 21]. However we present the *first* poly-logaritmic approximation algorithm for the general node-capacities case. The GROUP STEINER TREE problem with *group-demands* and *node-capacities* also generalizes other covering problems where the cover is required to be connected in some graph, like Connected dominating set problem or Connected set cover with hard capacities problem.

### 1.4 UNBALANCED POINT TO POINT CONNECTION

We next define a very important special case of CAP-SNDP called the UNBALANCED POINT TO POINT CONNECTION (UNBALANCED-P2P) problem. This problem is motivated by a so-called *Shift scheduling* problem with several practical applications to workforce scheduling. A solution to the shift design problem has been included in a product called OPA from Ximes Gmbh. [1]. See [13, 29] for more details.

> UNBALANCED POINT TO POINT CONNECTION (UNBALANCED-P2P)
> *Instance:* An undirected graph $G = (V, E)$ with edge-costs $\{c_e \mid e \in E\}$ and integer *charges* $\{b_v : v \in V\}$.
> *Objective:* Find a minimum-cost subgraph $H$ of $G$ such that $b(H') := \sum_{v \in H'} b_v \geq 0$ for every connected component $H'$ of $H$.

It is easy to see that the problem has a feasible solution if, and only if, $G$ is a feasible solution, i.e., every connected component $C$ of $G$ satisfies $b(C) \geq 0$, and that any inclusion-minimal solution is a forest. Given an instance of UNBALANCED-P2P, let $V^+ = \{v \in V \mid b_v > 0\}$ and let $V^- = \{v \in V \mid b_v < 0\}$. The fact that UNBALANCED-P2P is a special case of CAP-SNDP, even in the case of single demand, can be seen as follows. Given an instance of UNBALANCED-P2P, create a graph $G'$ by adding to $G$ two new nodes $s$ and $t$ and edges $(s, v)$ for all $v \in V^-$ and $(v, t)$ for $v \in V^+$. The original edges in $G$ inherit their cost $c_e$ and get infinite capacity. The new edges $(s, v)$ for $v \in V^-$ get capacity $|b_v|$ and zero cost and the new edges $(v, t)$ for $v \in V^+$ get capacity $b_v$ and zero cost. The nodes $s, t$ for the source-sink pair with connectivity requirement $|\sum_{v \in V^-} b_v|$.

Our result for UNBALANCED-P2P is as follows.

**Theorem 1.4** *There exists a polynomial-time combinatorial 2-approximation algorithm for the special case of* UNBALANCED-P2P *with* $b(V) := \sum_{v \in V} b_v = 0$. *Furthermore, if the charges* $\{b_v : v \in V\}$ *are polynomially bounded in* $|V|$, UNBALANCED-P2P *admits an exact algorithm on trees instances (i.e., $G$ is a tree) and ratio $O(\log \min\{n', 2 + b(V)\})$ on general graphs, where $n' = |V^+ \cup V^-|$ is the number of nodes with non-zero charge.*

Apart from being very important in practice, UNBALANCED-P2P generalizes the following important problems.

- Point-to-Point Connection [19]. This problem is exactly to a special case of UNBALANCED-P2P with $b_v \in \{-1, 0, 1\}$ for all $v \in V$ and $|V^+| = |V^-|$, i.e., $\sum_{v \in V} b_v = 0$. Goemans and Williamson [19] give a $(2 - \frac{1}{|V^-|})$-approximation algorithm for this problem.

- $k$-Steiner Tree [12]. The instance of this problem is given by an undirected graph $G = (V, E)$ with edge-costs $\{c_e \mid e \in E\}$, a subset $U \subseteq V$ of terminals and an integer $k \leq |U|$ and the goal is to find a minimum-cost tree in $G$ that contains at least $k$ terminals. The case $U = V$ is the $k$-MST problem [16]. The $k$-Steiner tree problem reduces to UNBALANCED-P2P as follows: "guess" a terminal $s$ that belongs to some optimal solution and set $b_s = -(k-1)$, $b_t = 1$ for all $t \in U \setminus \{s\}$, and $b_v = 0$ otherwise.

- Steiner Forest problem [19]. The instance of this problem is given by an undirected graph $G = (V, E)$ with edge-costs $\{c_e \mid e \in E\}$ and $k$ pairs of terminals $s_1 t_1, \ldots, s_k t_k$ and the goal is to find a minimum-cost subgraph of $G$ that connects $s_i$ to $t_i$ for all $i \in [k]$. Without loss of generality, we can assume that these pairs of terminals do not share a node. This problem reduces to UNBALANCED-P2P as follows: for $i \in [k]$, set $b_{s_i} = 2^i$, $b_{t_i} = -2^i$, and $b_v = 0$ otherwise. We argue that any feasible solution to this instance of UNBALANCED-P2P connects $s_i$ to $t_i$ for all $i \in [k]$ and vice versa. Since $\sum_{v \in V} b_v = 0$, each connected component in a feasible solution must have total charge zero. Thus the total positive charge (written out in the binary representation) must equal the absolute value of total negative charge (written out in the binary representation) in any connected component. Thus any connected component contains $s_i$ if and only if it contains $t_i$ for any $i \in [k]$.

## 1.5 Previous work

The CAP-SNDP is one of the most fundamental problems in combinatorial optimization. Even the Fixed-Cost Flow problem (the case of a single source and single sink) already includes several fundamental problems. Krumke et al. [28] proved a logarithmic hardness of the directed version, and gave a $k$-approximation algorithm, where $k$ is the requirement of the single pair. The special case of directed CAP-SNDP namely, directed Fixed-Cost Flow was shown to be Label-Cover hard by Even et al. [13] in 2002, which implies the same lower bound for CAP-SNDP. Eight years later, the same hardness was rediscovered independently by Chekuri et al. [8].

Goemans et al. [19] are the first who consider approximation algorithms for CAP-SNDP with multiple pairs. However they mainly consider "soft capacities", where multiple copies of an edge are allowed. Carr et al. [7] observed that the natural cut-based LP-relaxation has an unbounded integrality gap even for the unicast case. Motivated by this observation they strengthened the basic cut-based LP by adding so called Knapsack-Cover inequalities. Using these inequalities, they obtained constant factor approximation algorithms for some special graph topologies. However, in the general case, the integrality gap of the basic cut-based LP enhanced by Knapsack-Cover inequalities is $\Theta(n^2)$. Very recently, Chekuri et al. [8] considered various special cases of CAP-SNDP. For soft capacities, they give an $O(\log k)$ upper bound where $k$ is the number if pairs with positive requirement and $O(\log n)$ approximation ratio for the case when $r_{ij}$ are equal for all $i, j \in V$. They also show $\Omega(\log \log n)$ hardness result for the case of soft capacities. They gave no hardness result for the hard capacity case, as in CAP-SNDP. Approximation ratios or hardness results for the soft capacities case do not extend to CAP-SNDP.

A related problem, that also generalizes the Survivable Network problem (but without capacities) is the Node-Weighted Survivable Network problem [26, 24]; in this problem the costs/weights are on the nodes, every edge has capacity 1 and cost 0, and the cost of a subgraph is the sum of the cost of its nodes. The best known ratio for this problem is $O(r_{\max} \log n)$ [26], where $r_{\max} = \max_{i,j \in V} r_{ij}$ is the maximum requirement.

Garg, Konjevod, and Ravi [17] present an $O(\log N \cdot \log k)$-approximation algorithm for Group Steiner Tree on tree where $k$ is the number of groups, and $N$ is the maximum size of a group. Zosin and Khuller [34] give an alternative primal-dual approximation algorithm that solves an exponential linear program, and has ratio $O(\log^2 n)$. The first combinatorial polylogarithmic algorithm is by Chekuri et al. [10], that used the recursive greedy technique (see [9, 27, 33]), to obtain ratio $O(\log^{2+\epsilon} n)$. All the above upper bounds are closed to the best possible as Halperin and Krauthgamer [22] give a lower bound of $\Omega(\log^{2-\epsilon} n)$ for any fixed $\epsilon$, unless NP has a quasi-polynomial-time Las Vegas algorithm.

Finally we list the best known approximation ratios for the other important special cases of Cap-SNDP. The best known ratio for $k$-MST is 2 [16] and for $k$-Steiner Tree is 4 [12] (one way to get ratio 4 is to apply metric completion and move to the graph induced by terminals, loosing a factor of 2, and then using the 2 approximation algorithm [16] for $k$-MST on the graph induced by the terminals). The best approximation factor for Steiner Tree is roughly 1.39 [6]. For Steiner Forest, Point-to-Point connection, and Survivable Network, the best known ratio is 2, see [2, 19, 23], respectively.

Even et al. [13] obtain $O(\log |\sum_{v \in V^-} b_v|)$-approximation algorithm for Unbalanced-P2P. Our $O(\log(2 + |\sum_{v \in V} b_v|))$-approximation algorithm result in Theorem 1.4 is incomparable.

### 1.6 Organization

We begin by proving Theorem 1.1 in Section 2. Our recursive-greedy algorithm for Submodular Tree Cover is given in Section 3 and algorithms for Unbalanced-P2P in Section 4.

## 2 Hardness of Connected Cap-SNDP (Theorem 1.1)

Given an instance $(G = (V, E), \{c_e \geq 0 \mid e \in E\}, r, \{S_1, \ldots, S_k\})$ of Group Steiner Tree, we construct an instance of Connected Cap-SNDP as follows (see Figure 1 for an illustration). For a positive integer $k$, let $[k] = \{1, \ldots, k\}$. Construct a graph $G_+ = (V_+, E_+)$ from $G$ by adding some new nodes and edges as follows. Let $V_+ = V \cup \{s\} \cup \{g_i \mid i \in [k]\}$ and $E_+ = E \cup F$ where $F = \{\{s, v\} \mid v \in \cup_{i \in [k]} S_i\} \cup \{\{v, g_i\} \mid v \in S_i, i \in [k]\} \cup \{\{g_i, r\} \mid i \in [k]\}$. Each edge $e \in E$ is assigned cost $c_e$ and capacity $u_e = \infty$. Each edge $e = \{s, v\}$ for $v \in \cup_i S_i$ is assigned cost $c_e = 0$ and capacity $u_e = |\{i \mid v \in S_i, i \in [k]\}|$, i.e., number of groups $v$ belongs to. Each edge $e = \{v, g_i\}$ for $v \in S_i, i \in [k]$ is assigned cost $c_e = 0$ and capacity $u_e = 1$. Each edge $e = \{g_i, r\}$ for $i \in [k]$ is assigned cost $c_e = 0$ and capacity $u_e = |S_i| - 1$, i.e., one less than the number of nodes in group $S_i$. Finally we set sink as $t = r$ and demand as $d = \sum_{i \in [k]} |S_i| = \sum_{v \in V} |\{i \mid v \in S_i, i \in [k]\}|$.

Now we show the following one-to-one correspondence between the feasible solutions of the original Group Steiner Tree and that of the created Connected Cap-SNDP instance.

**Lemma 2.1** *There exists a solution for the Group Steiner Tree with cost at most $C$ if, and only if, there exists a solution for Connected Cap-SNDP instance with cost at most $C$. Furthermore, the solution to Group Steiner Tree can be computed in polynomial time from that to Connected Cap-SNDP instance, and vice versa.*

Let subtree $T = (V_T, E_T)$ be a solution of cost $C$ to the Group Steiner Tree instance. Let $H = E_T \cup F$ be a subgraph of $G_+$. Since all edges in $F$ have cost 0, the cost of $H$ is also $C$. We now argue that $H$ forms a feasible solution to the Connected Cap-SNDP instance, i.e., a flow of $d$ units can be routed from $s$ to $t$ in $H$. We start by routing flow of $u_{\{s,v\}} = |\{i \mid v \in S_i, i \in [k]\}|$ units from $s$ to each $v \in \cup_i S_i$. Consider a node $v \in V_T \cap (\cup_i S_i)$.

Such a node forwards its entire flow to $t = r$ along the unique path from it to $r$ along the tree $T$. This flow can be supported since the edges in $T$ have infinite capacity. Now consider a node $v \in (\cup_i S_i) \setminus V_T$. Such a node forwards 1 unit of its received flow to each $g_i$ for which $v \in S_i$ along the unit-capacity edge $\{v, g_i\}$. Note that any node $g_i$ receives at most $|S_i| - 1$ units of flow from all the nodes $v \in S_i$. This is because at most $|S_i| - 1$ nodes in $S_i$ do not belong to $T$, which in turn holds because $T$ contains at least one node from $S_i$. Lastly each node $g_i$ forwards its received flow to $t = r$ along edge $\{g_i, r\}$ of capacity $|S_i| - 1$. Thus indeed $H$ forms a feasible solution to the CONNECTED CAP-SNDP instance.

Now let $H$ be a solution of cost $C$ to the CONNECTED CAP-SNDP instance. Since all edges in $F$ have zero cost, we can assume that $F \subset H$, without loss of generality. It is enough to prove that $H \cap E$ contains a path from some node in $S_i$ to $r$ for each $i \in [k]$. Suppose this is not true for some group $S_j$ for $j \in [k]$. We extract an $s$-$t$-cut in graph $H$ with capacity strictly less than $d$ contradicting the existence of flow of value $d$ from $s$ to $t$ in $H$. Let $U \subset V$ denote the set of nodes connected to some node in $S_j$ in $H \cap E$ and let $\mathcal{U} = \{s, g_j\} \cup U$. Note that $s \in \mathcal{U}$ while from our assumption $t \notin \mathcal{U}$. We now prove the following claim.

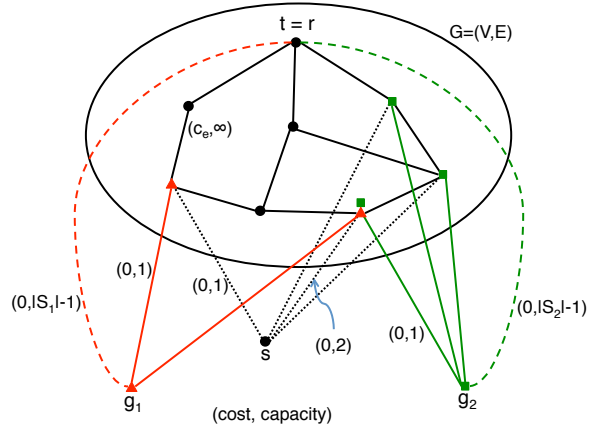

Figure 1: The instance of CONNECTED CAP-SNDP created in the reduction from GROUP STEINER TREE. The labels on the edges denote (cost, capacity). Not all labels are shown in the figure.

**Claim 2.2** *The total capacity of edges in $H$ that leave $\mathcal{U}$ is strictly less than $d$.*

*Proof:* It is easy to note that all the edges in $H$ that leave $\mathcal{U}$ are (1) $\{g_j, r\}$ with capacity $|S_j| - 1$, (2) $\{v, g_i\}$ with capacity 1, for all $i \neq j$ and $v \in S_i \cap U$, and (3) $\{s, v\}$ with capacity $|\{i \mid v \in S_i, i \in [k]\}|$ for all $v \in V \setminus U$. Thus the total capacity of these edges is

$$|S_j| - 1 + \sum_{i \neq j} \sum_{v \in S_i \cap U} 1 + \sum_{v \in V \setminus U} |\{i \mid v \in S_i, i \in [k]\}|$$
$$= |S_j| - 1 + \sum_{v \in U} |\{i \mid v \in S_i, i \in [k], i \neq j\}| + \sum_{v \in V \setminus U} |\{i \mid v \in S_i, i \in [k]\}|$$
$$= -1 + \sum_{v \in U} |\{i \mid v \in S_i, i \in [k]\}| + \sum_{v \in V \setminus U} |\{i \mid v \in S_i, i \in [k]\}|$$
$$= -1 + \sum_{v \in V} |\{i \mid v \in S_i, i \in [k]\}|$$
$$= d - 1.$$

The claim follows. ∎

The above claim implies Lemma 2.1, and thus the proof of Theorem 1.1 is complete.

# 3 Approximating SUBMODULAR TREE COVER (Theorem 1.3)

## 3.1 Preliminaries and notations

Given a set $U$, a function $f : 2^U \to \mathbb{Z}$, a subset $S \subseteq U$ and an element $x \in U$, denote $f_S(x) = f(S \cup \{x\}) - f(S)$ and $f_S(T) = f(S \cup T) - f(S)$. We say that $f$ obeys the *improvement independence* axiom if for every pair of subsets $S, T \subseteq U$ such that $S \subseteq T$, $\sum_{u \in T \setminus S} f_S(u) \geq f(T) - f(S)$. We recall the following equivalence from [32]: an increasing function $f : 2^U \to \mathbb{Z}$ is submodular if and only if it obeys the improvement independence axiom.

Let $f : 2^U \to \mathbb{Z}$ be a non-decreasing submodular function. By subtracting $f(\emptyset)$ from all values of $f$, we assume without loss of generality that $f(\emptyset) = 0$. Thus $f(A) \geq 0$ for all $A \subseteq U$. For any two subsets $A, B \subseteq U$, since $f(A \cap B) \geq 0$, the submodularity of $f$ implies $f(A) + f(B) \geq f(A \cup B)$.

We probabilistically embed the given graph into a tree metric losing $O(\log n)$ factor in the approximation, by using the results of Bartal [5] and Fakcharoenphol, Rao and Talwar [14]. There is a one-to-one correspondence between the original vertices $V$ and the set $L$ of leaves of a single embedding $T$. Using standard techniques, we also assume, without loss of generality, that we are solving the problem on a rooted tree instance where the root is required to be included in the output tree. We also assume, without loss of generality, that all leaves of $T$ are at the same level, i.e., level $h(T)$ where $h(T)$ denotes the height of tree $T$.

The parent of a non-root node $v$ is denoted by $p(v)$. The subtree rooted at a node $v$ is denoted by $T_v$. Let $e = (u, v)$ be an edge where $u$ is the parent of $v$. The subtree induced by the edge $(u, v)$ is the tree $T_v \cup \{(u, v)\}$, namely, the tree $T_v$ in addition to the edge $(u, v)$ and the node $u$. We denote the subtree induced by the edge $(u, v)$ by $T_{(u,v)}$. Let $n$ be the number of nodes in $T$.

The algorithm is recursive. In a general step of the algorithm, we have a tree $\tilde{T}$ that is to be included in the solution and we are computing an augmentation tree $T'$ to satisfy some demand. We abuse the notation and for a tree $T'$, use $f(T')$ to denote $f(L(T'))$ where $L(T') \subseteq L$ denotes the set of leaves included in $T'$. Similarly, we use $f_{\tilde{T}}(T')$ to denote $f(L(\tilde{T} \cup T')) - f(L(\tilde{T}))$, i.e., the increase in $f$-value due to the addition of $T'$ to $\tilde{T}$.

Let $\text{den}_{\tilde{T}}(T') = c(T')/f_{\tilde{T}}(T')$ denote the *density*, or cost to profit ratio, for subtree $T'$. Here $c(T') = \sum_{e \in T'} c_e$ denotes the cost of the tree $T'$. From submodularity of $f$, we get that for a collection of trees $\{T_i\}_{i=1}^p$,

$$\sum_{i=1}^p f_{\tilde{T}}(T_i) \geq f_{\tilde{T}}\left(\bigcup_{i=1}^p T_i\right). \tag{1}$$

## 3.2 Intuition

The algorithm and analysis is more complicated than that of the combinatorial group Steiner tree algorithm of Chekuri et al. [10]. Some of the complications are pointed out below. Some steps we take are:

- Using submodularity of $f$, we show that one can ignore subtrees of the optimum with low $f$-value. The proof is different (more detailed) than the one in [10].

- We have to guess, given some root $r$, the extent by which the tree rooted at $r$ in OPT, increases the current $f$-value. For efficiency reasons, we cannot check all values, and so we search in powers of roughly $1 + 1/\log n$. The fact that we dont search on all values creates a problem, the solution of which will become evident when the algorithm is given.

- We use the fact that we never make recursive calls with very small increase in $f$-value (because we ignore "small" trees) and we use geometric search on the amount of increase, to show the polynomial the running time.

- We are not increasing the number of terminals as in [10], but the value of $f$. In [10] as the terminals were "new", it was clear that those increases are independent. In our case, if we add trees $T_1, T_2, \ldots, T_p$ in this order, we need to show that the increase in $f$-value is large even after the previous trees were added, namely we need to show that $\sum_{j=1}^{p} f_{\tilde{T} \cup (\cup_{i=1}^{j-1} T_i)}(T_j)$ is large. The proof is more complicated than one in [10].

- Loosely speaking, we call the algorithm with target $z$ of increase in $f$-value and a tree of height $h'$. We stop at a critical point when the increase in $f$-value is at least $z/h'$. The density of this tree is used in the analysis. This is a point at which the density of the optimum solution has not change much yet. The original optimum was a candidate for addition by the algorithm in all previous iterations and its density is not much worse than the density of the optimum with respect to the empty set. We get a telescopic product that shows that the density derived is about $O(h')$ times the optimum density.

## 3.3 Height and degree reductions

We first recall the height and degree reductions from Chekuri et al. [10].

**Claim 3.1** [10] *There exists a combinatorial linear time algorithm that, given an instance of* SUBMODULAR TREE COVER *on a rooted tree $T$ with $\ell$ leaves, achieves the following. For an integer parameter $\alpha$, computes an instance $T'$ of* SUBMODULAR TREE COVER *such that the height of $T'$ is $O(\log_\alpha \ell)$ and for feasible solution $S$ for $T$ there exists a feasible solution $S'$ for $T'$ so that $c(S') \leq O(\alpha) \cdot c(S)$, and for every feasible solution $S'$ of $T'$, a feasible solution $S$ of $T$ can be computed in linear time such that $c(S) \leq c(S')$.*

**Claim 3.2** [10] *There exists a combinatorial linear time algorithm that, given an instance of the* SUBMODULAR TREE COVER *on a rooted tree $T$ with $\ell$ leaves and an integer parameter $\beta \geq 3$, computes a rooted tree $T'$ with height $h(T) + \lceil \log_{\beta/2} n \rceil$ such that every node has at most $\beta$ children. Moreover, for every feasible solution $S'$ for $T'$, there exists a feasible solution $S$ for $T$ with the same weight, and vice versa.*

For some $\epsilon > 0$, we set $\alpha = \beta = \log^\epsilon n$ and assume that the height is at most $O(\log_{\log^\epsilon n} \log n) = O(\log n / \epsilon \log \log n)$, maximum degree is $O(\log^\epsilon n)$, and the penalty in the approximation ratio is $O(\log^\epsilon n)$. For a node $v$, let $\deg(v)$ denote the number of children of $v$.

## 3.4 Ignoring small trees and geometric search

In a general step of the algorithm, suppose $\tilde{T}$ is the tree already included in the solution. To simplify the notation in the rest of this subsection, we assume that $\tilde{T} = \emptyset$ and update the definition of $f$ accordingly, i.e., we use $f(T')$ to denote $f_{\tilde{T}}(T')$ and $\texttt{den}(T')$ to denote $\texttt{den}_{\tilde{T}}(T')$. Note that such a change does not affect non-negativity, monotonicity and submodularity of $f$. Suppose our current target is to find an augmentation tree $T'$, rooted at some vertex $r$, so that $f(T') \geq z$ where $z$ is the target increase amount. The vertex $r$ and the augmentation amount $z$ are fixed throughout this subsection. Let $T^*$ be the minimum cost tree so that $f(T^*) \geq z$.

For a child $u$ of $r$, let $T^*_{(r,u)} = T^* \cap T_{(r,u)}$ denote the subtree of $T^*$ that hangs from the edge $(r, u)$. We let $\lambda = 1/h(T)$, where $h(T) = O(\log n / \log \log n)$ is the height of the entire tree $T$. We call $T^*_{(r,u)}$ *small* if $f(T^*_{(r,u_i)}) < \frac{z}{\deg(r) \cdot (1 + 1/\lambda)}$; and *big* otherwise. Let $\mathcal{F}$ be the forest of small trees. Let $T^{\text{big}} = T^* \setminus \mathcal{F}$. We now show that the density of $T^{\text{big}}$ is not much larger than the density of $T^*$.

**Lemma 3.3 (Ignore small trees)** $\texttt{den}(T^{big}) \leq (1 + \lambda) \cdot \texttt{den}(T^*)$.

*Proof:* Let $T_1, \ldots, T_p$ be the trees in $\mathcal{F}$. Using inequality (1) we get

$$f(T^*) \le f(T^{\text{big}}) + \sum_{i=1}^{p} f(T_i) \le f(T^{\text{big}}) + \deg(r) \cdot \frac{z}{\deg(r) \cdot (1 + 1/\lambda)} = f(T^{\text{big}}) + \frac{z}{(1 + 1/\lambda)}.$$

Therefore

$$f(T^{\text{big}}) \ge f(T^*) - \frac{z}{(1 + 1/\lambda)} \ge f(T^*)\left(1 - \frac{1}{(1 + 1/\lambda)}\right) = f(T^*) \cdot \left(\frac{1}{1 + \lambda}\right).$$

Thus, $\text{den}(T^{\text{big}}) = c(T^{\text{big}})/f(T^{\text{big}}) \le (1 + \lambda) \cdot c(T^*)/f(T^*) = (1 + \lambda) \cdot \text{den}(T^*)$, as desired. ∎

Since the density does not increase significantly, we are safe to ignore small trees. To increase $f$-value by $z$, we make several recursive calls with increments of $f$-value that are powers of $1 + \lambda$ in the range

$$\left[\frac{z}{\deg(r) \cdot (1 + 1/\lambda) \cdot (1 + \lambda)}, z\right].$$

Note that the lower end of this range is factor $1 + \lambda$ smaller than the term used in the definition of small trees. We restrict the search to powers of $1 + \lambda$ in order to ensure polynomial running time.

### 3.5 Greedy augmentation algorithm

Our algorithm is called `Greedy-Augment`. See Figure 2. The parameters are the vertex $r$ and value $z > 0$ and the goal is to find a tree rooted at $r$ that augments the $f$-value by at least $z$. We add trees one by one. The union of trees added so far is denoted by $\mathcal{C}$. As more trees are incorporated to $\mathcal{C}$, $f(\mathcal{C})$ gets larger and larger. The output of `Greedy-Augment` however may not end up augmenting $f$-value by at least $z$. If the height $h(T_r)$ of the tree $T_r$ rooted at $r$ is 1, we output a single edge. Otherwise, we make recursive calls and keep augmenting $\mathcal{C}$ till $f(\mathcal{C})$ is at lease at least $z$. Let $\mathcal{C}_h$ be the value of $\mathcal{C}$ when we have $f(\mathcal{C}) \ge z/h(T_r)$ for the first time. We eventually output the best density tree among $\mathcal{C}$ and $\mathcal{C}_h$.

The following lemma bounds the running time of the algorithm `Greedy-Augment`.

**Lemma 3.4** *[10] Let $\Delta$ be the maximum degree of the tree $T_r$ and let $\beta = \Delta(1 + 1/\lambda)(1 + \lambda)$. The algorithm `Greedy-Augment`$(r, z)$ takes $O(n\alpha^{h(T_r)})$ time and oracle calls to value oracle for $f$. Here $\alpha = \beta \cdot h(T_r) \cdot \log z \cdot \Delta \cdot \log_{1+\lambda} \beta$. If $h(T_r) = O(\log n / \log \log n)$, $\Delta = O(\log n)$, $1 \le 1/\lambda = O(\log n)$, $z$ is polynomially bounded in $n$ and if value oracle for $f$ takes time polynomial in $n$, then the overall running time is polynomial in $n$.*

The proof of this lemma is similar to [10] and is omitted. The value oracle for submodular functions $f$ needed for applications in Section 1.3 can be reduced to max-flow or min-cut algorithms.

We next prove the approximation guarantees of this algorithm.

**Lemma 3.5** *The output $T_{out}$ of `Greedy-Augment` satisfies*

$$\text{den}(T_{out}) \le (1 + \lambda)^{2h(T_r)} \cdot h(T_r) \cdot \text{den}(T^*).$$

#### 3.5.1 Proof of Lemma 3.5

The rest of this subsection is devoted to proving Lemma 3.5. The proof is by induction on the height of $T_r$. For base case, $h(T_r) = 1$, we note that the optimum augmentation tree $T^*$ is a star and the output consists of a

> Algorithm Greedy-Augment$(r, z)$ :
>
> 1. **Initialize:** $\mathcal{C} \leftarrow \emptyset$, $Z \leftarrow z$, and $T_{res} \leftarrow T_r$.
> 2. **Base case:** If $h(T_r) = 1$, **return** the edge $(r, u)$ where $u$ is a child of $r$ with minimum density $\text{den}((r, u)) = c((r, u))/f(u)$.
> 3. **While** $Z > 0$ **do:**
>    (a) **Recurse:** For every child $u$ of $r$ and for every $z'$ that is a power of $(1+\lambda)$ in $[\frac{Z}{\deg(r) \cdot (1+1/\lambda) \cdot (1+\lambda)}, Z]$, let $\mathcal{C}_{u,z'} \leftarrow$ Greedy-Augment$(u, z')$.
>    (b) **Select:** Let $T_{aug} \leftarrow \arg\min \text{den}(\mathcal{C}_{u,z'} \cup \{(r, u)\})$ be the minimum density tree among those computed.
>    (c) **Update:**
>       i. $\mathcal{C} \leftarrow \mathcal{C} \cup T_{aug}$.
>       ii. Update $Z$ as: $Z \leftarrow Z - f(T_{aug})$.
>       iii. Update function $f$ as: let $f(T')$ denote $f_{\tilde{T} \cup \mathcal{C}}(T')$.
>       iv. **If** it is first time $f(\mathcal{C}) \geq z/h(T_r)$, **then** $\mathcal{C}_h \leftarrow \mathcal{C}$.
> 4. **Return** lower density tree among $\mathcal{C}$ and $\mathcal{C}_h$.

Figure 2: The Greedy-Augment algorithm for SUBMODULAR TREE COVER

single edge $(r, u^*)$. By submodularity of $f$, we have

$$\text{den}(T^*) = \frac{c(T^*)}{f(T^*)} \geq \frac{\sum_{(r,u) \in (T^*)} c((r,u))}{\sum_{(r,u) \in T^*} f(u)} \geq \min_{(r,u) \in T^*} \frac{c((r,u))}{f(u)} = \frac{c((r,u^*))}{f(u^*)} = \text{den}((r, u^*)).$$

Now we prove the induction step. The proof here is different than one in [10]. Recall that $T^{\text{big}}$ is the union of big trees in $T^*$. Decompose $T^{\text{big}}$ into the trees $T^*_{(r,u_1)} \cup T^*_{(r,u_2)} \cup \cdots \cup T^*_{(r,u_k)}$. Here tree $T^*_{(r,u_i)}$ is a tree $T^*_{u_i}$ rooted at child $u_i$ of $r$ plus the edge $(r, u_i)$. Say that tree number $i$ is rooted by $u_i$. Let $z^*_i = f(T^*_{(r,u_i)})$. By a simple averaging argument, it follows that the density of at least one of the big subtrees is at most $\text{den}(T^{\text{big}})$. Without loss of generality, assume that

$$\text{den}(T^*_{(r,u_1)}) \leq \text{den}(T^{\text{big}}) \leq (1+\lambda) \cdot \text{den}(T^*). \qquad (2)$$

Let $z_1 = (1+\lambda)^i$ be such that $z_1 \leq z^*_1 < (1+\lambda) \cdot z_1$. Note that $z_1$ is in the range of powers of $(1+\lambda)$ considered in line 3a in Algorithm Greedy-Augment. Here we see why the least value of the search interval needs to be the term used to define small trees divided by $1+\lambda$. We upper bound the density by considering a very specific recursive call. Then we can bound the density by the induction hypothesis. Consider the call $\mathcal{C}_{u_1, z_1} \leftarrow$ Greedy-Augment$(u_1, z_1)$ in line 3a of Greedy-Augment. We now upper bound the density for that call. The tree $C_{u_1, z_1}$ is incrementally constructed from a sequence of augmenting trees, denoted by $\{R_1, R_2, \ldots\}$. Let $j$ denote the smallest integer such that

$$f\left(\bigcup_{i=1}^{j} R_i\right) \geq \frac{z_1}{h(T_r)}. \qquad (3)$$

Note that when computing $R_j$, the $f$-value of the union is less than $z_1/h(T_r)$. During all the iterations of the while loop in which $C_{u_1,z_1}$ is computed, the subtree $T^*_{u_1}$ is a feasible solution for the required increase in $f$-value to $z_1$. Thus by definition, for $p \leq j - 1$,

$$f\left(\bigcup_{i=1}^{p} R_i\right) \leq \frac{z_1}{h(T_r)}. \qquad (4)$$

Since $f$ is non-decreasing and submodular,

$$z_1 \leq z_1^* = f(T_{u_1}^*) \leq f\left(\left(\bigcup_{i=1}^{p} R_i\right) \cup T_{u_1}^*\right) \leq f\left(\bigcup_{i=1}^{p} R_i\right) + f_{\bigcup_{i=1}^{p} R_i}(T_{u_1}^*).$$

Plugging in the inequality (4) we get

$$f_{\bigcup_{i=1}^{p} R_i}(T_{u_i}^*) \geq z_1 - \frac{z_1}{h(T_r)}.$$

Thus

$$\text{den}_{\bigcup_{i=1}^{p} R_i}(T_{u_1}^*) \leq \frac{c(T_{u_1}^*)}{z_1 - z_1/h(T_r)} \tag{5}$$

The following establishes an upper bound on $\text{den}_{\bigcup_{i=1}^{p} R_i}(R_{p+1})$.

**Claim 3.6** *For all $p \leq j - 1$,*

$$\text{den}_{\bigcup_{i=1}^{p} R_i}(R_{p+1}) \leq (1+\lambda)^{2h(T_r)-2} \cdot h(T_r) \cdot \frac{c(T_{u_1}^*)}{z_1}.$$

*Proof:* By Inequality (5) and the induction hypothesis we get:

$$\begin{aligned}
\text{den}_{\bigcup_{i=1}^{p} R_i}(R_{p+1}) &\leq (1+\lambda)^{2h(T_{u_1})} \cdot h(T_{u_1}) \cdot \frac{c(T_{u_1}^*)}{z_1 - z_1/h(T_r)} \\
&= (1+\lambda)^{2h(T_r)-2} \cdot (h(T_r) - 1) \cdot \frac{c(T_{u_1}^*)}{z_1 - z_1/h(T_r)} \\
&= (1+\lambda)^{2h(T_r)-2} \cdot h(T_r) \cdot \frac{c(T_{u_1}^*)}{z_1}.
\end{aligned}$$

∎

Here is a claim that is needed only as $f$ is submodular.

**Claim 3.7**

$$\text{den}(\mathcal{C}_h) \leq (1+\lambda)^{2h(T_r)-2} \cdot h(T_r) \cdot \frac{c(T_{u_1}^*)}{z_1}.$$

*Proof:* Note that $\mathcal{C}_h = R_1 \cup R_2 \cdots \cup R_j$. Now

$$\begin{aligned}
\text{den}(\mathcal{C}_h) = \frac{c(\mathcal{C}_h)}{f(\mathcal{C}_h)} &= \frac{\sum_{p=1}^{j} c(R_p)}{\sum_{p=1}^{j} f_{\bigcup_{i=1}^{p-1} R_i}(R_p)} \\
&\leq \max_{1 \leq p \leq j} \text{den}_{\bigcup_{i=1}^{p-1} R_i}(R_p) \\
&\leq (1+\lambda)^{2h(T_r)-2} \cdot h(T_r) \cdot \frac{c(T_{u_1}^*)}{z_1}.
\end{aligned}$$

∎

Finally, we bound the density of $T_{out}$.

$$\begin{aligned}
\text{den}(T_{out}) &\leq \frac{c(\cup_{i \leq j} R_i) + c_{(r,u_1)}}{f(\cup_{i \leq j} R_i)} \\
\text{(by Claim 3.7)} &\leq (1+\lambda)^{2h(T_r)-2} \cdot h(T_r) \cdot \frac{c(T^*_{u_1})}{z_1} + \frac{c_{(r,u_1)}}{z_1/h(T_r)} \\
\text{(by def. of } z_1^*) &\leq (1+\lambda)^{2h(T_r)-2} \cdot h(T_r) \cdot \frac{c(T^*_{u_1})}{z_1^*/(1+\lambda)} + h(T_r) \cdot \frac{c_{(r,u_1)}}{z_1^*/(1+\lambda)} \\
&\leq (1+\lambda)^{2h(T_r)-1} \cdot h(T_r) \cdot \frac{c(T^*_{u_1}) + c_{(r,u_1)}}{z_1^*} \\
\text{(by definition)} &= (1+\lambda)^{2h(T_r)-1} \cdot h(T_r) \cdot \text{den}(T^*_{(r,u_1)}) \\
\text{(by Eq. 2)} &\leq (1+\lambda)^{2h(T_r)} \cdot h(T_r) \cdot \text{den}(T^*).
\end{aligned}$$

This proves Lemma 3.5.

## 3.6 Putting things together

We run `Greedy-Augment` iteratively till we obtain a tree with the maximum value $F_{\max}$ of $f$. By a simple set-cover like argument, the overall running time is polynomial in $n$ times $\log F_{max}$ and the overall approximation ratio for the tree-instances $T$ is $O(\log^\epsilon n \cdot h(T) \cdot \log F_{\max}) = O(\log^{1+\epsilon} n \cdot \log F_{\max})$. The first $\log^\epsilon n$ factor comes due to height and degree reductions. For the graph instances, we get $O(\log^{2+\epsilon} n \cdot \log F_{\max})$-approximation, where another $\log n$ term comes due to approximating the general metric by tree metrices.

# 4 Approximating GEN-P2P CONNECTION (Theorem 1.4)

## 4.1 A 2-approximation algorithm for the case $b(V) = 0$

Our 2-approximation algorithm is an easy extension of the algorithm of [19, 18] for the Point-to-Point Connection problem, which is the case $b_v \in \{-1, 0, 1\}$. We say that an edge $e$ covers a set $S$ if $e$ has exactly one endnode in $S$; we say that an edge-set/graph covers a set family $\mathcal{F}$ if for every $S \in \mathcal{F}$ there is an edge in $H$ covering $S$. Given a set-family $\mathcal{F}$ and an edge-set $H$ the residual set-family $\mathcal{F}_H$ consists of the members of $\mathcal{F}$ not covered by $H$. Recall that a set-family $\mathcal{F}$ is *uncrossable* if for any $X, Y \in \mathcal{F}$ at least one of the following holds: $X \cap Y, X \cup Y \in \mathcal{F}$ or $X \setminus Y, Y \setminus X \in \mathcal{F}$. It is known and easy to see that if $\mathcal{F}$ is uncrossable, so is $\mathcal{F}_H$, for any edge-set $H$.

Goemans et al. [18] give a primal-dual 2-approximation algorithm for the problem of finding a minimum-cost edge-cover of an uncrossable set-family $\mathcal{F}$. A polynomial time implementation of this algorithm requires only that for any edge-set $H$, the minimal members of the residual set-family $\mathcal{F}_H$ can be computed in polynomial time (but $\mathcal{F}$ itself may not be given explicitly). Now the 2-approximation algorithm follows from the following lemma.

**Lemma 4.1** *Given an instance of* UNBALANCED-P2P *with $b(V) = 0$, let $\mathcal{F} = \{S \subseteq V \mid b(S) \neq 0\}$. Then the following holds.*

(i) *An edge-set $H \subseteq E$ is a feasible solution to* UNBALANCED-P2P *if, and only if, $H$ covers $\mathcal{F}$.*

(ii) *For any edge set $H \subseteq E$, $S$ is an inclusion-minimal members of $\mathcal{F}_H$ if, and only if $S$ is a connected component of the graph $(V, H)$ and $b(S) \neq 0$.*

(iii) $\mathcal{F}$ *is uncrossable.*

*Proof:* Parts (i) and (ii) are straightforward, so we prove only part (iii). Let $X, Y \in \mathcal{F}$, so $b(X), b(Y) \neq \emptyset$. We will show that if $X \cap Y \notin \mathcal{F}$ or if $X \cup Y \notin \mathcal{F}$, then $X \setminus Y, Y \setminus X \in \mathcal{F}$. Suppose that $X \cap Y \notin \mathcal{F}$, so $b(X \cap Y) = 0$. Then $b(X \setminus Y) = b(X) - b(X \cap Y) = b(X) \neq 0$ and $b(Y \setminus X) = b(Y) - b(Y \cap X) = b(Y) \neq 0$; hence $X \setminus Y, Y \setminus X \in \mathcal{F}$. Suppose that $X \cup Y \notin \mathcal{F}$, so $b(X \cup Y) = 0$. Then $b(X \setminus Y) = b(X \cup Y) - b(Y) = -b(Y) \neq 0$ and $b(Y \setminus X) = b(X \cup Y) - b(X) = -b(X) \neq 0$; hence $X \setminus Y, Y \setminus X \in \mathcal{F}$. ∎

## 4.2 An exact algorithm for trees

We now focus on the case when the charges $b_v$ are polynomially bounded, but the total charge $b(V)$ may not be zero. We show how to solve the problem on trees optimally, using dynamic programming.

Root the tree $T$ at some node $s$. By adding zero-cost edges to $T$ if necessary, we can assume that $T$ is a binary tree without loss of generality. In particular, if a node $v$ has $p$ children, we add a binary tree with $p$ leaves at $v$ and connect $p$ leaves one-to-one to the $p$ leaves. We give a cost of zero to each of the tree edges. It is easy to see that the instance essentially remains unchanged by this modification. For a node $v \in T$, let $T_v$ denote the subtree hanging below $v$. The dynamic program computes quantities $\mathcal{T}(v, B)$ for all nodes $v \in T$ and integer $B$ in the range $[\sum_{u:b_u<0} b_u, \sum_{u:b_u>0} b_u]$. Since each $b_u$ is polynomially bounded, the number of such quantities is polynomial. The quantity $\mathcal{T}(v, B)$ is defined as the minimum-cost of a subgraph $H$ of $T_v$ satisfying the following:

- the connected component in $H$ containing $v$ has the total charge $B$, and
- every other connected component in $H$ has non-negative total charge.

If there is no subgraph $H$ satisfying the above conditions, we define $\mathcal{T}(v, B)$ as $-\infty$. We assume that the minimum-cost subgraph $H$ is also stored in the dynamic program table.

The quantities $\mathcal{T}(v, B)$ can be computed as follows. For leaf nodes $v$, it is trivial to compute $\mathcal{T}(v, B)$ and the corresponding optimum subgraphs. For an internal node $v$, we compute $\mathcal{T}(v, B)$ as follows. Let $u_1$ and $u_2$ be the two children of $v$. Depending on whether we pick edges $(v, u_1)$ or $(v, u_2)$ in the solution, we get four possibilities.

1. If we pick none of these edges in the solution, we get a solution of cost $\min\{\mathcal{T}(u_1, B_1) + \mathcal{T}(u_2, B_2) \mid B_1, B_2 \geq 0\}$ corresponding to charge of the connected component containing $v$ of $b_v$.

2. If we pick edge $(v, u_1)$ but do not pick edge $(v, u_2)$ in the solution, we get a solution of cost $\min\{c_{(v,u_1)} + \mathcal{T}(u_1, B_1) + \mathcal{T}(u_2, B_2) \mid B_2 \geq 0\}$ corresponding to charge of the connected component containing $v$ of $b_v + B_1$.

3. If we pick edge $(v, u_2)$ but do not pick edge $(v, u_1)$ in the solution, we get a solution of cost $\min\{c_{(v,u_2)} + \mathcal{T}(u_2, B_2) + \mathcal{T}(u_1, B_1) \mid B_1 \geq 0\}$ corresponding to charge of the connected component containing $v$ of $b_v + B_2$.

4. If we pick both the edges $(v, u_1)$ and $(v, u_2)$ in the solution, we get a solution of cost $\min\{c_{(v,u_1)} + \mathcal{T}(u_1, B_1) + c_{(v,u_2)} + \mathcal{T}(u_2, B_2)\}$ corresponding to charge of the connected component containing $v$ of $b_v + B_1 + B_2$.

We consider all these possibilities and pick the minimum-cost solution corresponding to each value of the charge of the connected component containing $v$.

Finally, we output the solution corresponding to $\min\{\mathcal{T}(s, B) \mid B \geq 0\}$. It is easy to see that the above dynamic programming based algorithm computes the optimum solution our problem.

## 4.3 An $O(\log n')$-approximation algorithm where $n' = |V^+ \cup V^-|$

The algorithm is as follows. We reduce the general problem to the case when the input graph is a tree with a loss of $O(\log n')$ factor in the approximation ratio. This is achieved as follows. Consider the shortest-path metric on $V' = V^+ \cup V^-$ w.r.t. the edge-costs $c_e$. We probabilistically embed this metric into a tree metric $T, c'$ with $O(\log n')$ distortion using the results of Bartal [5] and Fakcharoenphol, Rao and Talwar [14]. There is a one-to-one correspondence between $V'$ and the set $L$ of leaves of $T$. The resulting instance of UNBALANCED-P2P on $T$ inherits the charges on the leaves of $T$ from the original charges on nodes of $V'$, while the charge of internal nodes of $T$ is 0. We compute an optimal solution to the obtained tree instance, and return the corresponding subgraph $H$ of $G$. Note that any feasible solution with cost $C$ for the original instance induces a solution with cost $O(C \log n')$ for the new instance on tree $T$. Similarly any feasible solution with cost $C$ for the new instance induces a solution with cost $C$ for the original instance. Hence the approximation ratio is bounded by the distortion of the reduction, which is $O(\log n')$.

Now consider the augmentation version of the problem, when we are give an edge subset $E' \subseteq E$ of cost 0. Then we can contract every connected component $F$ of $(V, E')$ into a single node $v_F$ with charge $b(v_F) = b(F)$. Thus the approximation ratio in this case is $O(\log n')$, where here $n'$ is the number of connected components with non-zero charge in the graph $(V, E')$.

## 4.4 An $O(\log(2 + b(V)))$-approximation algorithm

Note that $b(V)$ may be very small (close to 0 even).

**Lemma 4.2** *There exists a polynomial time algorithm that given an instance of* UNBALANCED-P2P *computes an edge set $E' \subseteq E$ of cost $\le 4\tau^*$, where $\tau^*$ denotes the optimal solution value, such that the number $n'$ of connected components with non-zero charge in the graph $(V, E')$ is at most $4b(V)$.*

*Proof:* Consider the following procedure that runs with a parameter $\tau$, which is an estimate for $\tau^*$. Create an instance of UNBALANCED-P2P with total charge zero by adding a new node $s$ with charge $-b(V)$ and connecting $s$ to each node in $V^+$ by an edge of cost $\tau/b(V)$. Then apply the 2-approximation algorithm for the case $b(V) = 0$. The new instance admits a solution of cost at most $\tau^* + b(V) \cdot (\tau/b(V)) = \tau^* + \tau$, by taking an optimal solution to the original instance with edges that connect $s$ to at most $b(V)$ nodes in $V^+$. Thus the procedure returns an edge-set of cost at most $2(\tau^* + \tau)$. Consequently, if $\tau \ge \tau^*$ then the procedure returns an edge-set of cost at most $4\tau$, and the number of edges incident to $s$ is at most $4\tau/(\tau/b(V)) = 4b(V)$. Using binary search, we find the minimum integer $\tau$ for which the procedure returns an edge-set $E''$ of cost $4\tau$. Then $c(E'') \le 4\tau \le 4\tau^*$ and the number of edges in $E''$ incident to $s$ is at most $4b(V)$. Let $E'$ be obtained from $E''$ by removing the edges incident to $s$. Then $c(E') \le c(E) \le 4\tau^*$, and the number $n'$ of connected components in $(V, E')$ with non-zero-charge is at most the degree of $s$ w.r.t. $E''$, hence at most $4b(V)$, as claimed. ∎

The entire algorithm has two steps. At step 1 we compute an edge set $E'$ as in Lemma 4.2. Step 2 applies the $O(\log n')$-approximation algorithm from the previous section to compute an augmenting edge-set $F \subseteq E \setminus E'$ such that $E' \cup F$ is a feasible solution. The solution cost is bounded by $c(E') + c(F) = O(\tau^*) + O(\log n') \cdot \tau^* = O(\log(2 + b(V))) \cdot \tau^*$.

## 5 Conclusions

We present hardness results and combinatorial algorithms for several special cases of CAP-SNDP. Naturally, obtaining a poly-logarithmic approximation algorithm for CAP-SNDP is a wide open question.

It is also open whether one can achieve a constant ratio for UNBALANCED-P2P. If so, it will be a single algorithm that gives a constant ratio for both Steiner Forest and $k$-Steiner Tree (or $k$-MST). Currently, constant ratio algorithms for these two problems use quite different algorithms. Thus a constant approximation algorithm for UNBALANCED-P2P, if possible, will unify techniques for both problems.